\documentclass[article]{aastex62}

\def\beq{\begin{equation}}
\def\eeq{\end{equation}}

\begin{document}

\title{Modeling stellar abundance patterns resulting from the addition of
earthlike planetary material}

\correspondingauthor{Charles R. Cowley}
\email{cowley@umich.edu}

\author{Charles R. Cowley}
\affiliation{Department of Astronomy, University of Michigan,
1085 S. University, Ann Arbor, MI 481090-1107}

\author{Donald J. Bord}
\affiliation{Department of Natural Sciences, University of Michigan-Dearborn,
4901 Evergreen Rd., Dearborn, MI 48128}

\author{Kutluay Y\"{u}ce}
\affiliation{Department of Astronomy and Space Sciences,
Faculty of Science \\ University of Ankara, Ankara, TR-06100, Turkey}

\keywords{stars: abundances --- sun: abundances --- planetary systems: planet-star interactions}
\begin{abstract}

We model the observed precision differential abundance
patterns of three G-type dwarfs, assuming a mass of
planetary or disk material was added to or subtracted from
the atmosphere of the star. The two-parameter model is fit
by linear regression. The stellar abundance patterns are
corrected for Galactic chemical evolution (GCE). The
differential abundances can be highly correlated with the
elemental condensation temperature. We show how it is
possible to match not only the slope, but the quantitative
abundance differences, assuming a composition for the added
material equal to that of the bulk earth. We
also model the wide pair HIP 34407 and HIP 34426, where GCE
corrections are assumed unnecessary. An appendix discusses
issues of volatility and condensation temperature.

\end{abstract}
\section{Introduction\label{sec:intro}}

The literature on precision differential abundances (PDAs)
in stars is
extensive.  Surveys include sun-like stars in the
solar neighborhood, binary systems, and Galactic clusters.
Numerous references as well as a discussion of
relevant mechanisms may be found in papers by \citet{ram19}
and \citet{nag20}.
A strong impetus for this work is the probability that
the abundances have been influenced by exoplanetary
systems and their evolution.

We calculate the resulting differential abundances
($[El/H]$) assuming a given amount of material with the composition
of the bulk earth \citep[][henceforth, WN18] {wang18} was added to the stellar
convection zone of a dwarf G-type star.  The mass of the convection
zone is uncertain and variable, depending on the spectral type.
Here, we assume a mass of $5\cdot 10^{31}$ gm
for the stellar convection zone (SCZ).  This is 0.025 $M_\odot$
\citep[cf.][]{pin01,cham10}.
For other SCZ masses, the parameters must be adjusted accordingly.
In general, the sunlike star will not have exactly the solar
composition.  This contingency is roughly taken into account in
our model.
\section{The model\label{sec:model}}
Let $M_{SCZ}$ be the mass of the stellar convection zone.
We use
a bulk earth mass of $M_{BE} = 6\cdot 10^{27}$ gm, with the composition
from WN18.
Our model assumes that as a result of the addition of $q$ earth masses of material,
the stellar abundance of all elements will be increased
by a factor of
\beq
\frac{q\cdot M_{BE}+p\cdot M_{SCZ}}{M_{SCZ}} = p + q\frac{M_{BE}}{M_{SCZ}}.
\label{eq:pandq}
\eeq
\noindent Note that the ratio of the masses, BE and SCZ, is the same as
the ratio of the number of {\em atoms} in the BE and the SCZ.

The parameter $p$ allows for the possibility that the star may
have abundances that differ slightly from those of the sun.  If $p < 1$,
the star will be slightly metal poor, and conversely, if $p > 1$, it will be
slightly metal rich.  Generally,
we expect that $p$ will be close to unity.  We assume
the abundances vary in lockstep, and neglect small changes in the
atmospheric structure resulting from the added material.

Using bracket notation, where $[X] = \log(X_{star})-\log(X_{sun}) = \log(X_{star}/X_{sun})$,
\beq
[El/H] = \log\left(p + q\frac{M_{BE}}{M_{SCZ}}\right).
\label{eq:elhpq}
\eeq
\noindent Rewriting Eq.~\ref{eq:elhpq} yields
\beq
p +  q\frac{M_{BE}}{M_{SCZ}} = 10^{[El/H]}.
\label{eq:tento}
\eeq
\noindent We may obtain $p$ and $q$ by linear regression, as there
is one equation for every differential elemental abundance.  The
regression fit includes the results of an $F$-test in which the $F$ statistic (the ratio
of the varance for a one parameter fit to the variance for a two parameter fit) is calculated
and used to establish a ``significance" based on the Snedecor $F$
distribution.   In this formalism, larger values of the $F$ statistic yield smaller significance measures;
outcomes with significances substantially less than 0.01 demonstrate the superiority of the
linear (two parameter) fit over that of flat line (one parameter or ``mean") fit.
Values for $p$, $q$, $F$, and the significance are included in the discussions of the examples
below.

\section{Examples}

\citet{adib14} pointed out that, in addition to correlations of differential abundances of solar-like
stars with condensation temperature, $T_{\rm c}$,
\footnote{Appendix A provides
a discussion of the condensation temperature.  In the present figures, unless otherwise noted,
the $T_{\rm c}$ values are from \citet{wood19}.}
the PDAs also correlated with the stars' ages and/or
locations in the disk.  These connections are readily seen as a consequence of
Galactic chemical evolution (GCE), whose overall effect is to convert mostly
light, volatile elements
like carbon, nitrogen, and oxygen into heavier, refractory elements like iron,
zirconium, and barium.  Thus, stars younger than the sun are expected to be depleted in
volatiles like C, N, and O but enhanced in refractories, such as
Fe, Zr, and Ba, and heavier elements.  This leads
to a positive correlation
of their differential abundances with condensation temperature.  Note that most
of the observable heavy elements have high $T_{\rm c}$.
There are, of course, exceptions to this rough trend.  Some heavy elements, such
lead and mercury, are volatile, and the heavier noble gasses are ultra volatile.
However abundance determinations for these elements are not usually available for G dwarfs.

It is necessary to subtract the effects of GCE prior to examining
the differential abundances for possible effects of addition or subtraction of
planetary material.  This was done by \citet{cby20a}\footnote{{see Table 1
only in arXiv:2003.14336}} in their study of
$T_{\rm c}$ correlations among the 79 stars of \citet[][henceforth, BD18]{bed18}
which included the first three examples below.  BD18 give precision
differential abundances for the following 30 elements: C, O, Na, Mg, Al, Si,
S, Ca, Sc, Ti, V, Cr, Mn, Fe, Co, Ni, Cu, Zn, Sr, Y, Zr, Ba, La, Ce, Pr, Nd,
Sm, Eu, Gd, and Dy.

In addition to GCE, the fascinating results of \citet{mel14} for 18 Sco show that the
abundances of neutron-addition (nA) elements (Sr-Dy in BD18) may be affected by local
nucleosynthetic events unrelated to general GCE.
In the current article we have not, therefore,  included the nA elements in the
fits of $[El/H]$ to $T_{\rm c}$ for HD 196390, HD 42618, and HD 140583.

\subsection{Data sources}
The calculations presented below require use of literature sources for condensation
temperature, the composition of the bulk earth, and the composition of the convection zone.  We
have made numerous calculations based on different combinations of data sources,
with the conclusion that for our purposes, differences in numerical values in the
data sources are not significant.

In the current article, we report calculations from models based on two sets of data
sources:
\begin{itemize}
\item Set I: Solar photospheric abundances from
\citet{ABA18,ABC19,AGG20,AGS09},
\citet{LAA17,BCA17,carl19,GSA15},
\citet{NL17,OYL19,RAL19},
\citet{SGA15,YNG18},
condensation temperatures from \citet{wood19}, and bulk earth composition from
\citet{wang18}.
\item Set II: Solar abundances from \citet[][see Table 6, Col 5]{lod20}.  These
abundances made partial use of CI determinations, as explained in the work cited.
Condensation temperatures were from \citet{lod03}, and bulk earth composition
from \citet{mcd08}.
\end{itemize}

We emphasize that for our purposes, which involve only the elements between
carbon and zinc, the choice of data sources is not critical.
As an illustration, we show in Tab.~\ref{tab:feg} that the derived model parameters
of Eq.~\ref{eq:tento}, $p$ and $q$, in data Sets I and II agree well.

\begin{deluxetable}{llrrrl}  [h]
\tablecaption{Model parameters from 2 data sets\label{tab:feg}}
\tablecolumns{6}
\tablehead{
\colhead{Set} &
\colhead{Star(s)} &
\colhead{p} &
\colhead{q} &
\colhead{F} &
\colhead{sig}
}
\startdata
Set I  &HD 196390     &1.07 &  5.0 & 54.0 & 1.4 E-07 \\
Set II &              &1.09 &  5.4 & 38.8 & 1.2 E-06\\ \hline
Set I  &HD 42618      &0.81 & -1.4 &  6.3 & 1.0E-02 \\
Set II &              &0.80 & -1.6 &  6.3 & 1.0E-02  \\ \hline
Set I  &HD 140538     &0.97 &  4.6 & 19.1 & 7.5 E-05 \\
Set II &              &1.00 &  4.4 &  8.3 & 3.7E-03 \\ \hline
Set I  &HIP34407/34426&1.28 & 8.2  & 19.7 & 2.9E-05  \\
Set II &              &1.31 & 6.6  & 14.9 & 1.5E-04
\enddata
\end{deluxetable}

\subsection{HD 196390 (HIP 101905)}
In \citet{cby20a}, six stars
had differential abundances with significant ($< 6.0\cdot 10^{-5}$)
$T_{\rm c}$ correlations;
the tightest correlation was for HD 196390.
\begin{figure}  [h]    
\epsscale{1.1}
\begin{center}
\plottwo{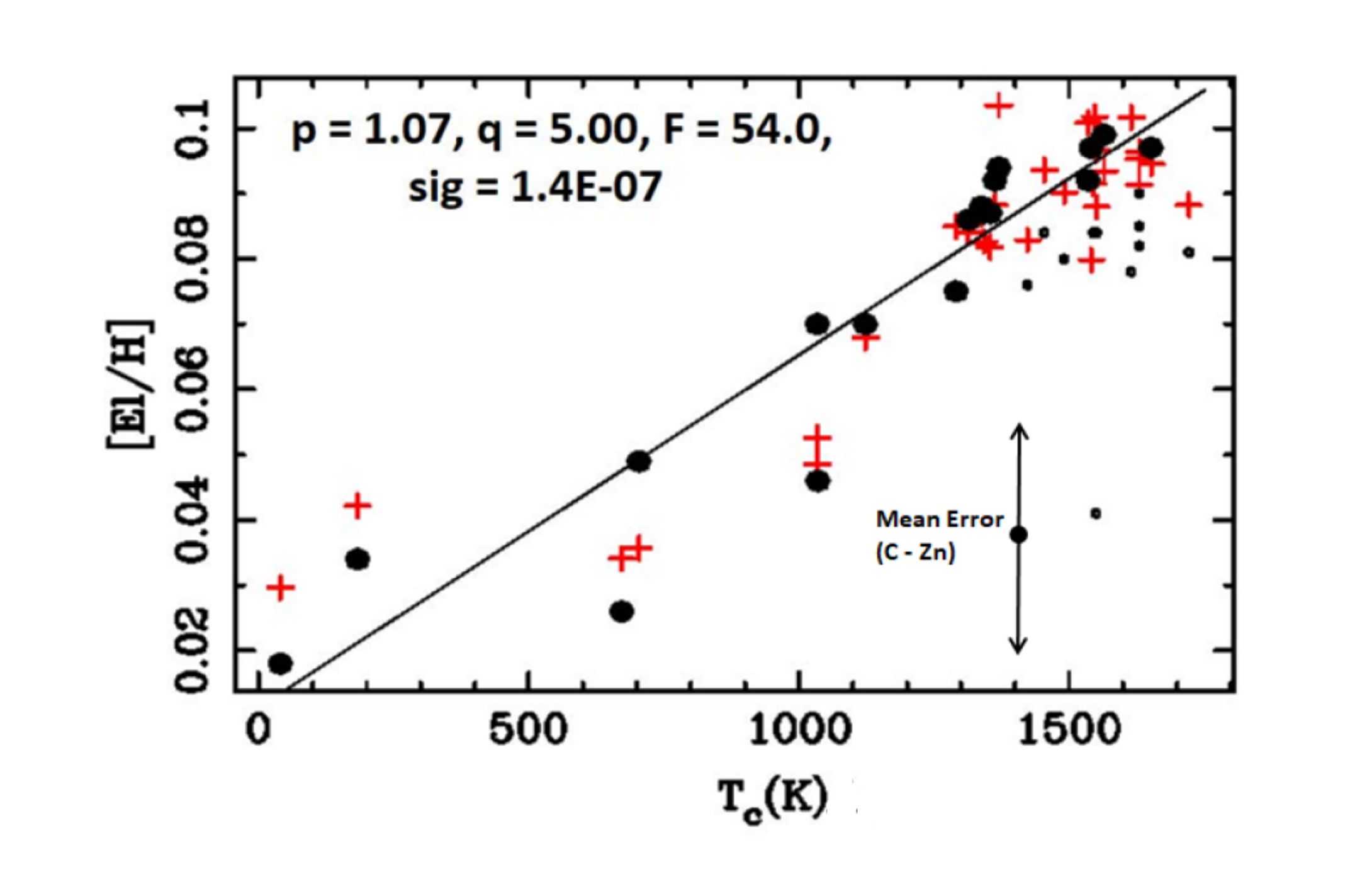}{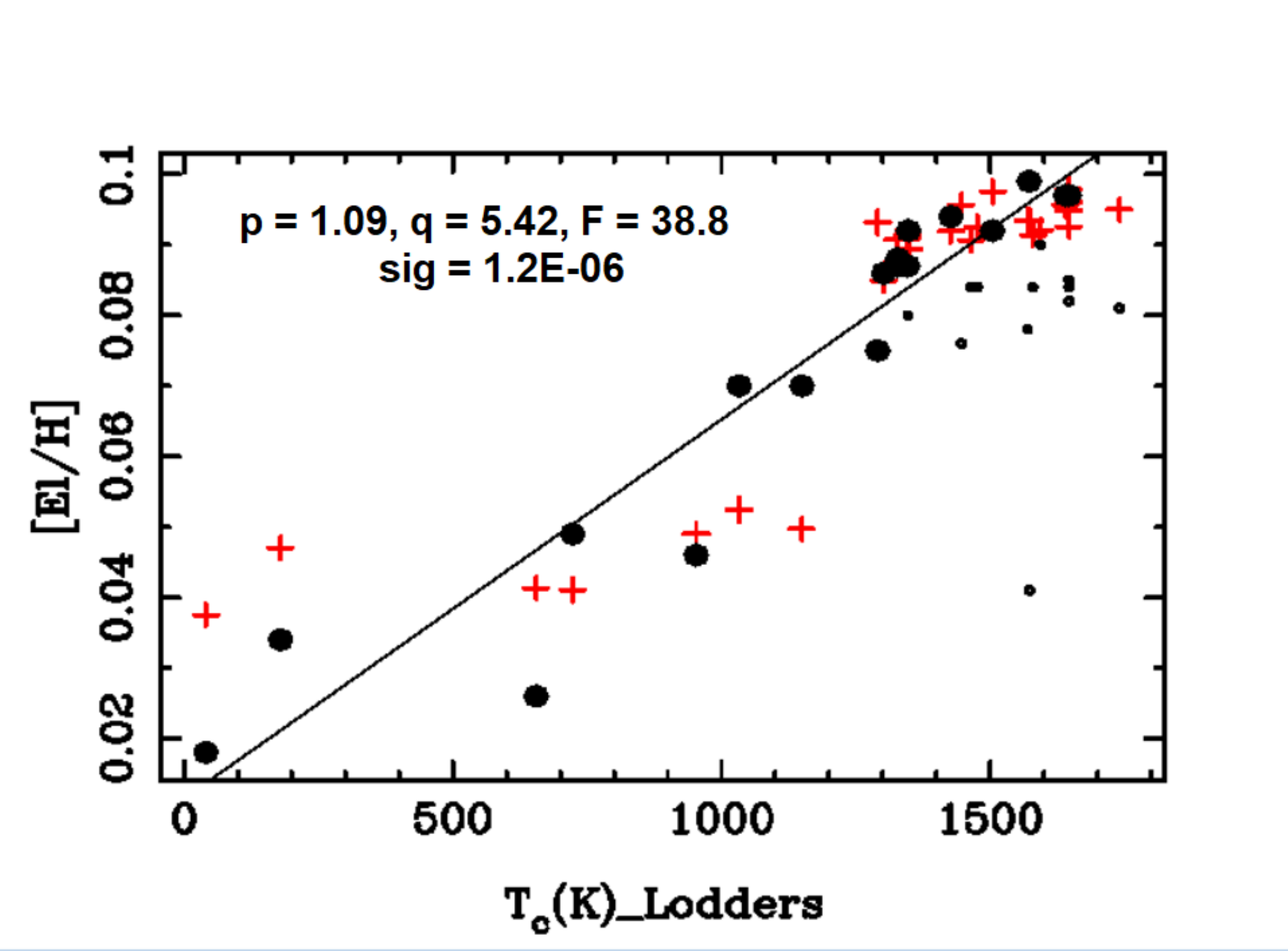}
\caption{Left: Model predictions (red crosses) and observed differential stellar abundances
(large black points) for HD 196390
vs. condensation temperature, $T_{\rm c}$, after GCE corrections.  The
parameters are from Set I of Tab.~\ref{tab:feg}.
The predicted abundances are from a slightly
metal-rich model ($p = 1.07$).  The parameter $q = 5.0$ implies
that five earth masses were ingested.  See text for further explanation
of the symbols.
\newline Right: Model predictions and observed differential stellar abundances
for HD 196390 based on the parameters of Set II of Tab.~\ref{tab:feg}.
Symbols have the same meaning as at left.  The two plots are closely similar.
The predicted point (red plus sign at $T_ c$ = 1158K) some
0.02 dex below the mean relation indicated by the solid black line is for Mn.
\label{fig:101}}
\end{center}
\end{figure}

Model fits are shown in
Fig.~\ref{fig:101} based on data Sets I and II of Tab.~\ref{tab:feg}.
The straight line is an unweighted least squares fit to $[El/H]$
vs. $T_{\rm c}$ for carbon through zinc.
The linear regression solution yielding $p$ and $q$ is also for these elements.
Diffferences between observations and models, for both Sets I and II, are typically
within the total uncertainties  of the abundance measures, including the observational
and analysis errors, plus those associated with the GCE correction parameters from BD18.
These errors are indicated on the left plot.
The small black dots designate the nA elements which were excluded from the
modeling trials and the regression fits.



\subsection{HD 42618 (HIP 29432)}

Among the stars of BD18 noted by \citet{cby20a}, HD 42618 was
unusual because of its negative slope, and because it has a known
planet containing some 14.4 earth masses.  The formal probability that an
unweighted linear fit of $[El/H]$ with $T_{\rm c}$ occurs by chance was
$2.1\cdot 10^{-5}$, but the two-parameter linear regression fit is only of marginal
significance.  Snedecor's F-parameter is 6.3, leading to a
significance level 0.010 (see Fig. 2, left).
Data Sets I and II yield $p < 1$  for this star, indicating
that HD 42618 is slightly metal poor with respect to the sun. The negative values of $q$
suggest a deficit of earthlike material relative to the sun.
\begin{figure} [h]   
\epsscale{1.1}
\begin{center}
\plottwo{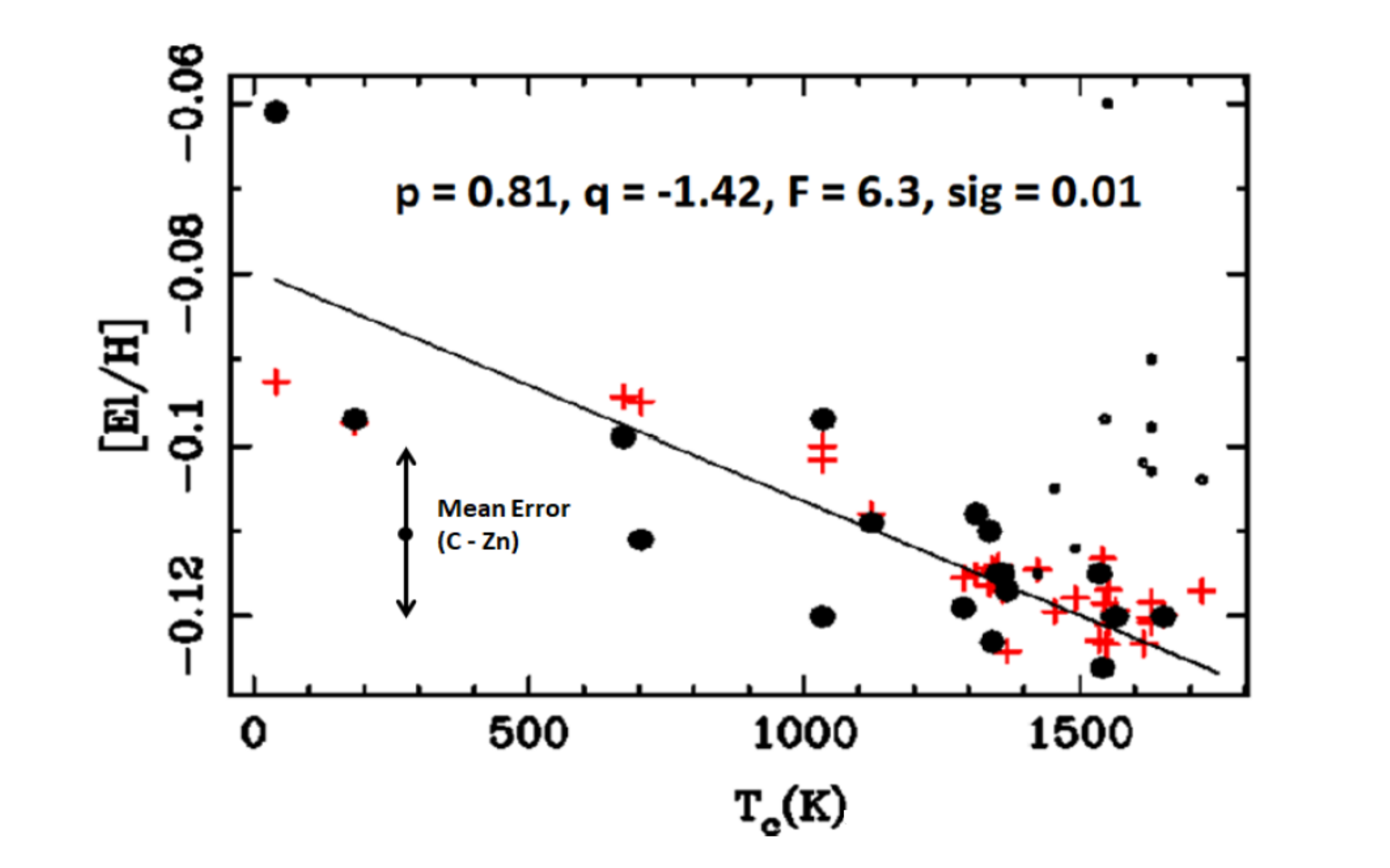}{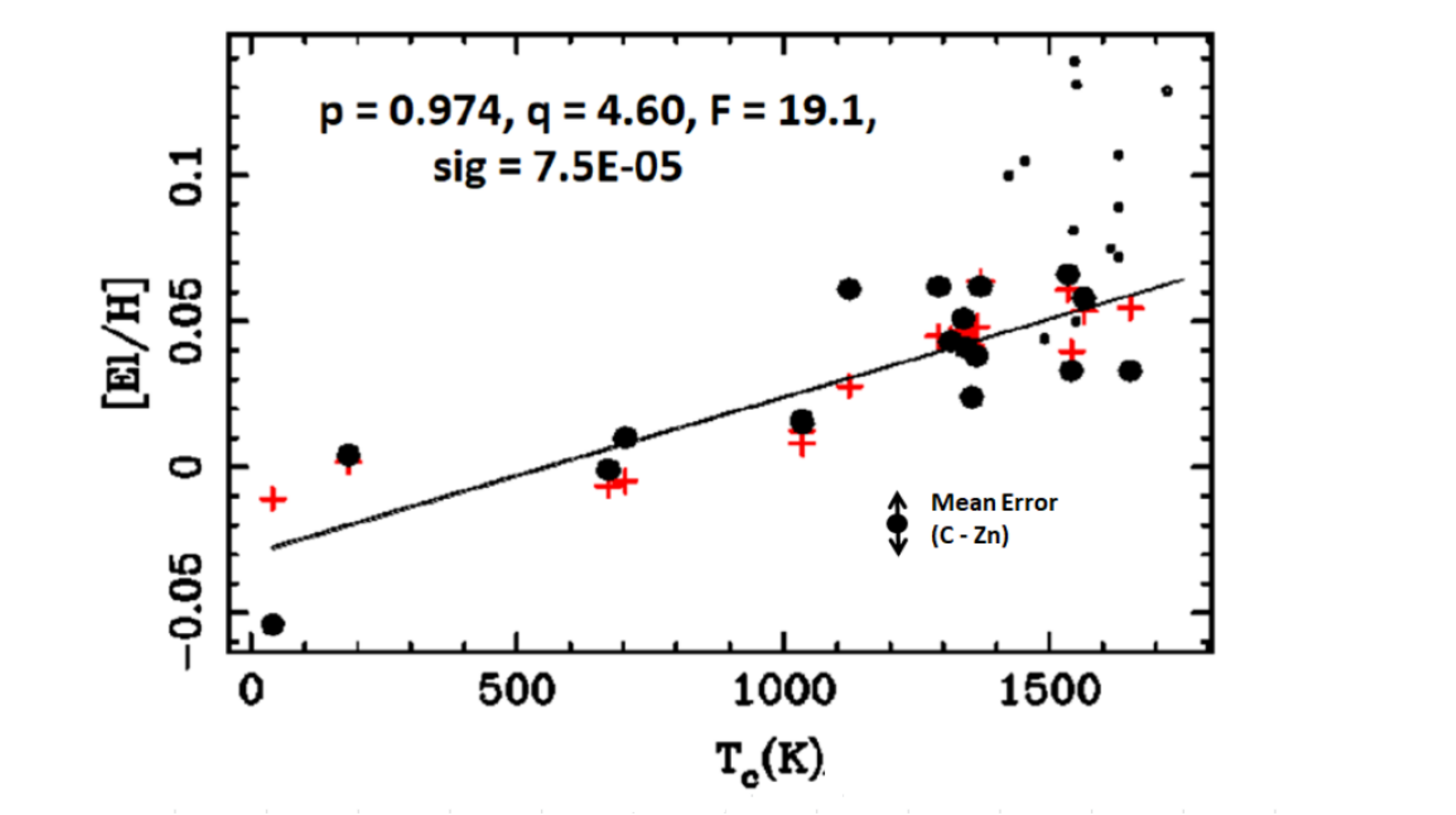}
\caption{Left: Model predictions and observed differential stellar abundances for HD 42618
vs. condensation temperature after GCE corrections.  The model abundances are slightly
metal poor.  The negative $q$-parameter implies that refractory mass
is missing from HD 42618 relative to the sun. \newline
Right: Model predictions and observed differential abundances for HD 140538 vs. condensation
temperature after GCE corrections. The model abundances are close to solar, with
$\approx 4.5$ earth masses of added material needed to optimize the fit to the observed PDAs.
\newline Both plots use data and results from Set I.}
\label{fig:618140}
\end{center}
\end{figure}

\subsection{HD 140538 (HIP 77052)}

HD 140538 ($\psi$ Ser, Fig. 2, Right) is among the stars of
\citet{cby20a} which have a significant
$[El/H]$ vs. $T_{\rm c}$ plot.  Although classified as G2.5 V, it is not considered
a solar twin possibly because it is thought to be in a triple system, with two
dwarf M companions \citep{mad16}.
The $p$-parameters for both Sets I and II are near unity, indicating solar
abundances for the 4 to 5 earth masses ($q$ parameter) added to HD 140538.



\subsection{The wide double, HIP 34407 and HIP 34426}

In the case of the binary pair, HIP 34407/HIP 34426, abundances are from \citet{ram19}
for 21 elements: Li, O, Na, Mg, Al, Si, S, K, Ca, Sc, Ti, V, Cr, Mn, Fe, Co,
Ni, Cu, Zn, Y, and Ba.  No GCE corrections were applied in this case, as we assume the stars
are coevil.  Moreover, if abundances for the two nA
elements {\em were} perturbed by some isolated nucleosynthetic event, we assume
both stars were equally influenced; thus, we include the nA elements Y and Ba
in the model fit, based on data Set I.

\begin{figure}
\epsscale{0.7}
\begin{center}
\plotone{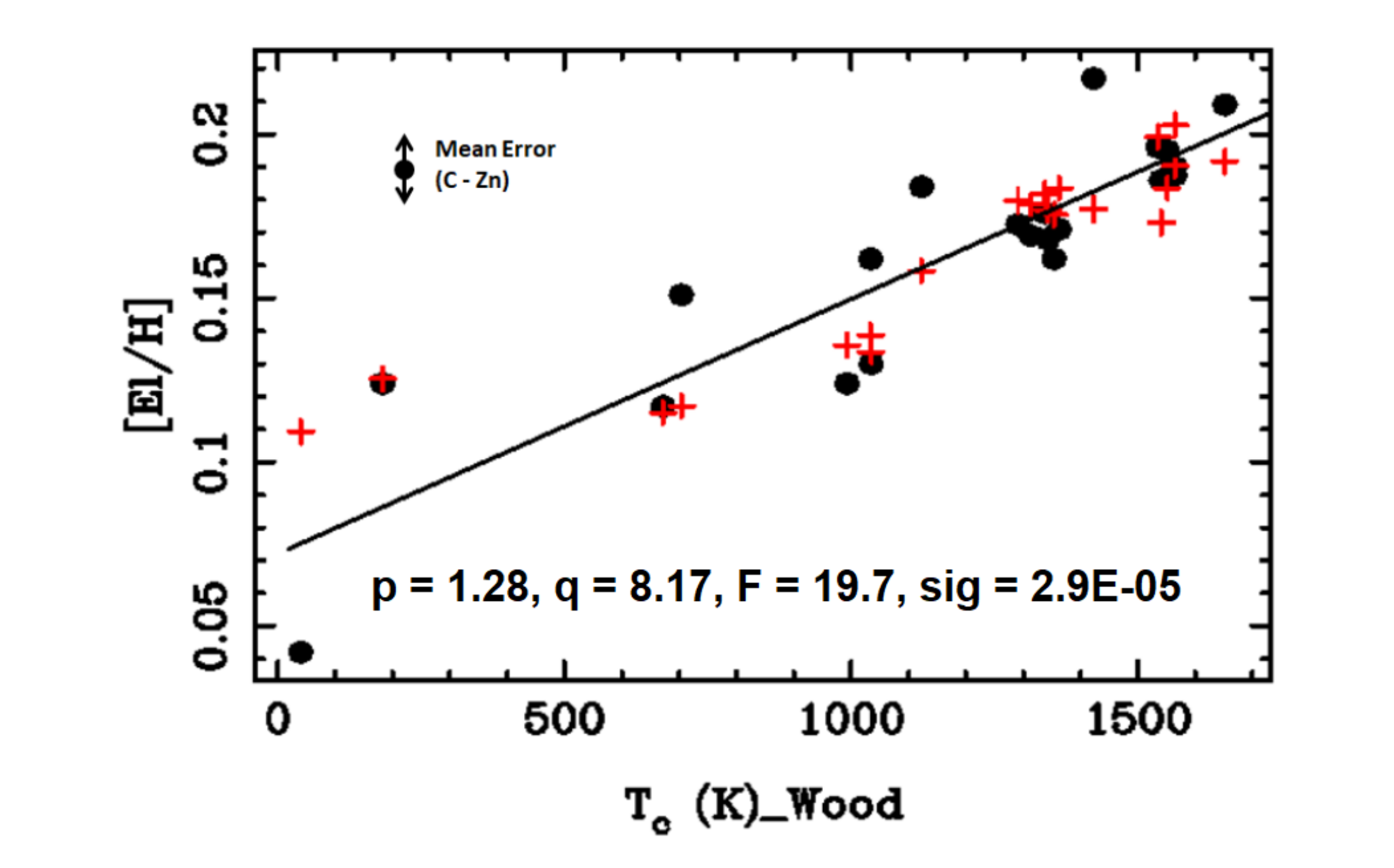}
\caption{
Model predictions and observed differential stellar abundances for HIP34407/HIP34426,
 where $[El/H]$ is $\log(El/H)_{\rm HIP 34407}-\log(El/H)_{\rm HIP 34426}$.
Symbols have the same meaning as in Fig.~\ref{fig:101}.  The model is based on
data and results of Set I.}
\label{fig:538}
\end{center}
\end{figure}

The comoving, closely similar G-stars, HIP 34407 and HIP 34426, were
part of a larger work
by  \citet{nag20}, who suggested the differential abundance pattern could
result from the engulfment of a gas giant.
The differential abundances are fit by addition of some 7 earth masses of material,
the largest of the modeled stellar masses of this work.  This is much lower than the total mass
of a Jupiter-sized gas giant, but not so different from estimates of the core mass of heavy
elements in Jupiter itself derived from models based on 1995
Galileo probe data (14 - 18 $M_E$, \citet{mil08})
and more recent results from the Juno mission (7 - 25 $M_E$, \citet{wah17}). We assume
that the more volatile elements forming the bulk of the mass of captured gas giants would be lost
during the engulfment process and not incorporated into the SCZ.

The lithium abundance in both stars is much higher than one would expect from
their ages, which are greater than 6.5 Gyr for all plausible models
according to \citet{ram19}, who give
$\log(Li/H)+12 = 2.37$ for HIP 34407 and 2.31 for HIP 34426.  Figs. 2-5 of \citet{carl19}
show that only the {\em youngest} stars in the 79-star BD18 sample have comparable values with
$\log(Li/H)+12 \approx 2.3$.  \citet{ram19}
discuss lithium abundances in both stars, but find no likely scenario to account for
either the high individual abundances or the small abundance difference between the two
stars. They compute
$\Delta [Li/H] = \log(Li/H)_{\rm HIP 34426}-\log(Li/H)_{\rm HIP 34407} = -0.05$;
this element is clearly above the trend for neighboring elements with similar
condensation temperatures
in their Fig. 4.  In our Fig.~\ref{fig:538}, in which
the roles of the two stars are reversed and
$[El/H]$ is $\log(El/H)_{\rm HIP 34407}-\log(El/H)_{\rm HIP 34426}$,
lithium would again be an obvious outlier in a plot
with $[Li/H] = 0.62$ at $T_{\rm c} = 1148$K.  Li is not included in the current model fit or plot.

  \section{Summary}

In this paper, we have shown that some G-type stellar PDAs can be meaningfully
interpreted by assuming that earth-like material was ingested by or missing
from the modeled stars, perhaps through planetary formation processes.
Four examples are presented in the form of plots of PDAs vs. $T_{\rm c}$,
three with positive slopes
and one with a negative slope. The derived model parameters confirm that the
mean composition of the SCZs of these stars is essentially solar ($p_{avg} =
1.03 \pm 0.20$) as expected, but show a range of mass adjustments from -1.4
to +8.2 bulk earths. Predicted model abundances typically replicate the
observed PDA distributions to within the observational errors which have
mean values for the elements C through Zn ranging from 0.011 to 0.023 dex,
including uncertainties associated with the parameters used to effect GCE
corrections. We have tested our approach using two different sets of source
data for the condensation temperatures, and the compositions for the bulk
earth material and the initial stellar convection zone and find that, for
the 30 elements included in our models, the results from the two sets are in
good agreement.

While linear fits like those shown in our Figures 1 - 3 are adequate to
describe the current results of PDA investigations
which have generally been limited to 30 or so well-studied elements, such
models would not properly describe $T_{\rm c}$ trends if precison abundance
data for the {\em full range} of naturally occuring elements in the periodic
table were available (cf. the Appendix). We argue that if the full
range of abundances were available, the slope in the low-$T_{\rm c}$, volatile
region should be essentially zero, while in the high-$T_{\rm c}$, refractory
regime, the slope should also approach zero, although with much greater
scatter.

\appendix

\section{Condensation temperatures--accuracy and relevance}
\begin{figure} [h]  
\begin{center}
\epsscale{0.6}
\plotone{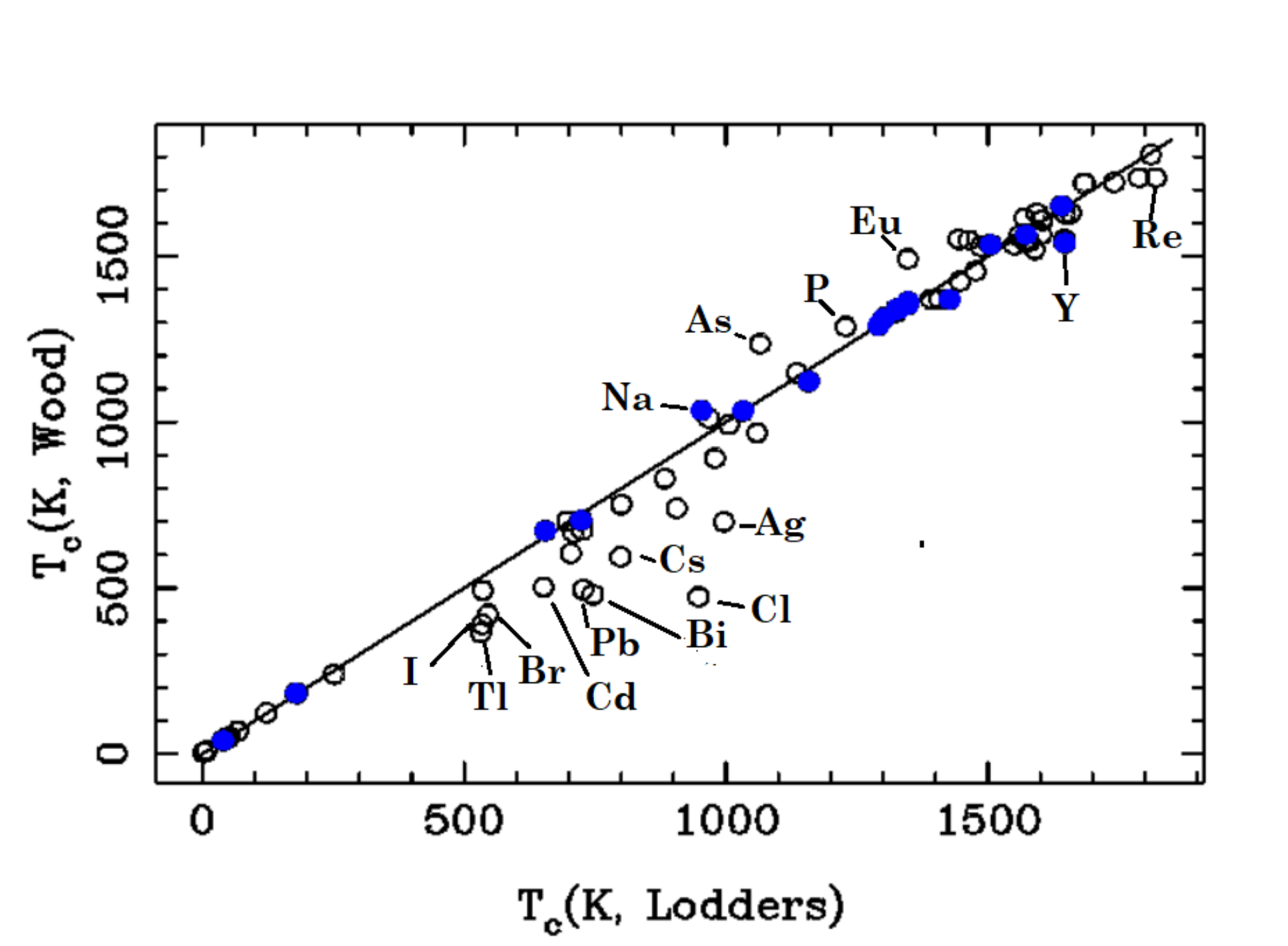}
\caption{Condensation temperatures from \citet{wood19} vs. those from
\citet{lod03}.
The highlighted points are for the 18 elements used for the present
models.  Of those elements, sodium falls slightly above the
50-50 line, while yttrium falls slightly below.  The latter element
was used only in the model for HIP 34407/34426.  Both deviations are
small relative to the uncertainties of the overall model calculations
presented in Sec. 3.}
\label{fig:tctc}
\end{center}
\end{figure}

The relevance of condensation temperatures to cosmic solids was established
by \citet{was74}.
The 50\% condensation temperatures of \citet{lod03}, and the more recent values by
\citet{wood19}, are in quite good agreement, apart
from a few elements, viz. Cl, Pb, As, and Bi none of which are included
in our modeling calculations (see Fig.~\ref{fig:tctc}).
The assumptions underlying the determinations
are more relevant.  Both calculations begin with a gas of solar
composition at a pressure of $10^{-4}$ bar,
and assume strict chemical equilibrium--gas-solid reactions
are allowed.
A detailed discussion of equilibrium calculations and
the geochemical classification of elements is given by \citet{feg10}.

\citet{bar76} discuss an alternate approach.
``At  the opposite extreme from the equilibrium model is the disequilibrium
condensation model, which prohibits reactions between gas and already
condensed phases, or between two or more condensed phases.''
Details of their calculations, available in
graphic form (see their Figs. 4 and 5), show qualitatively similar
results.  These figures also show that changes in the gas pressure
from $10^{-2}$ to $10^{-6}$ bar do not significantly change the order of
condensation.  A useful recent discussion is by \citet{li20} who give
2\% condensation temperatures,
also for a solar mix and a pressure of $10^{-4}$ bar, based on the GRAINS code
of \citet{pet09}.  In all these calculations, volatile elements remain volatile
and refractories refractory.
\begin{deluxetable}{lrrrrr|c}  [h]
\tablecaption{Bulk earth: representative element abundance estimates\label{tab:1}}
\tablecolumns{7}
\tablewidth{0pt}
\tablehead{
\colhead{Element} &
\colhead{Mc03} &
\colhead{Mc08} &
\colhead{WN18}&
\colhead{\% Error}&
\colhead{50\% $T_{\rm c}$(K)} &
\colhead{VM}
}
\startdata
C &   730  & 730    &   2 648&84  &40  &0.00469    \\
N &    25  & 57     &     31 &87  &123 &0.000196   \\
O &297 000 &297 000 &308 000 &2.3 &183 &0.229  \\   \tableline
Zn&    40  &35      &   45.9 &8.7 &704 &0.113     \\
S &  6 350 &6 300   &   6 096&24  &672 &0.084     \\
Cu&    60  &60      &     69 &2.8 &1034&0.420     \\
Na&  1 800 &1 800   &   2 201&8.5 &1035&0.345     \\  \tableline
Fe&320 000 &319 000 & 312 000&0.3 &1338&108   \\
Mg&154 000 &153 000 & 151 000&1.3 &1343&1.00
\enddata
\end{deluxetable}

Tab.~\ref{tab:1} shows three estimates of the bulk earth abundance,
for representative highly volatile, intermediate volatile, and
refractory elements.  Horizontal lines divide the three categories
of volatility.
Abundances are in parts per million by mass, from
\citet[Mc03]{mcd03}, \citet[Mc08]{mcd08}, and WN18.
The
fifth column gives WN18's estimate of their errors in per cent.
The condensation temperatures are from \citet{wood19}.  The final column
is a measure of the element's volatility (VM).  It
gives the ratio:
\begin{equation}
VM =
\left[\frac{\rm mass\ of\ element\ in\ earth}{\rm mass\ of\ magnesium\ in\ earth}\right] \div
\left[\frac{\rm mass\ of\ element\ in\ sun}{\rm mass\ of\ magnesium\ in\ sun}\right]
\label{eq:fracfrac}
\end{equation}
Tab.~\ref{tab:1} illustrates the persistance of the property of volatility
for different choices of earth models.  Note that oxygen has an abundance that
would put it with the involatiles, in spite of its low $T_{\rm c}$.  This, of
course, is because of the high reactivity of oxygen, so that it combines to 
form the ubiquitous silicates. 

We conclude that statistically significant fits of $[El/H]$ data as a function
of 50\% $T_{\rm c}$, whether strictly linear, piece-wise linear,
or logarithmic \citep[cf.][]{wang19}, are robust indicators of the
condensation of cosmic material, earthlike planets, or meteorites.
The slopes of the relevant relations may
be positive, zero or negative.

\subsection{Insights from the full abundance table\label{sec:details}}

\begin{figure} [h]   
\epsscale{1.1}
\begin{center}
\plottwo{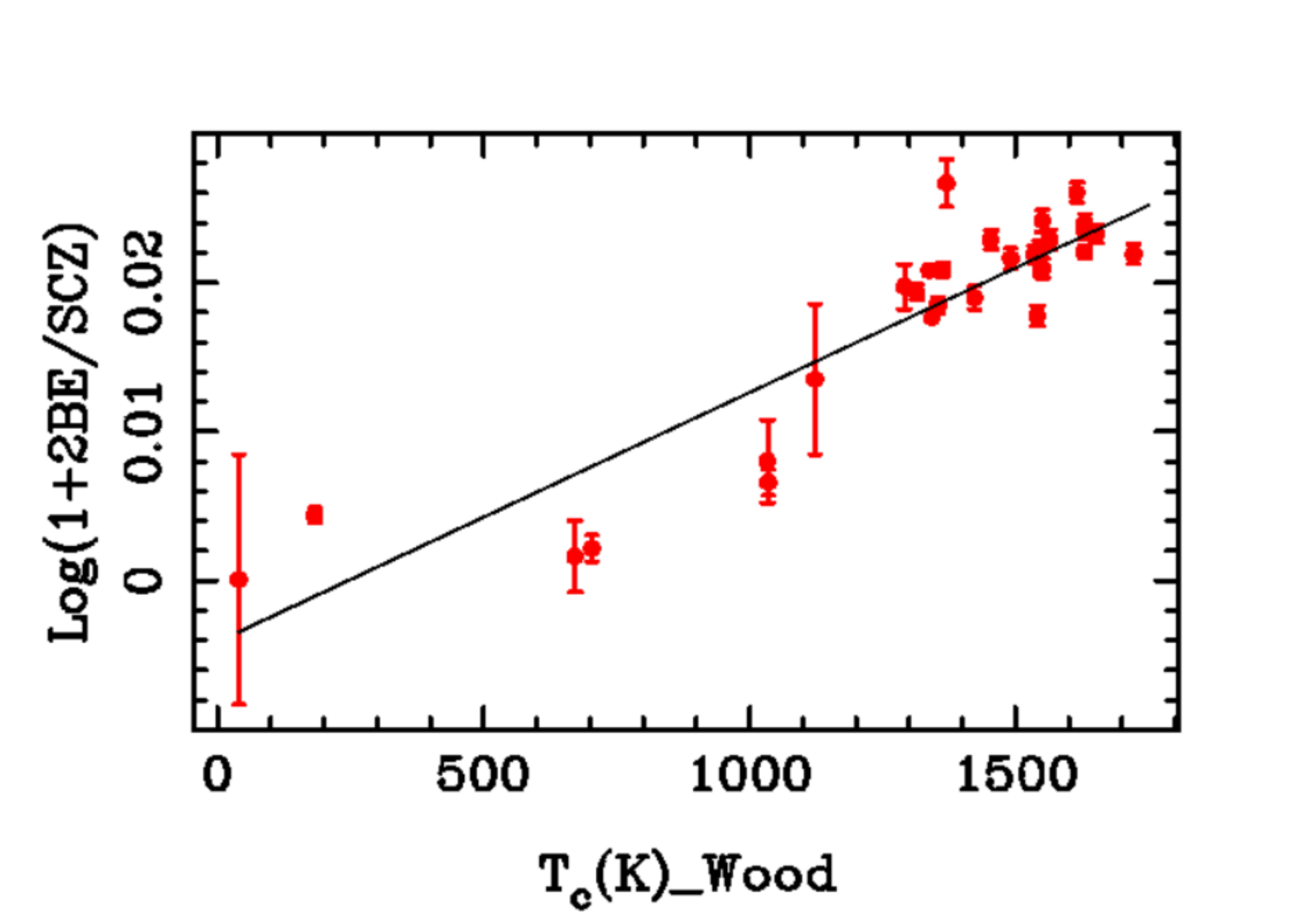}{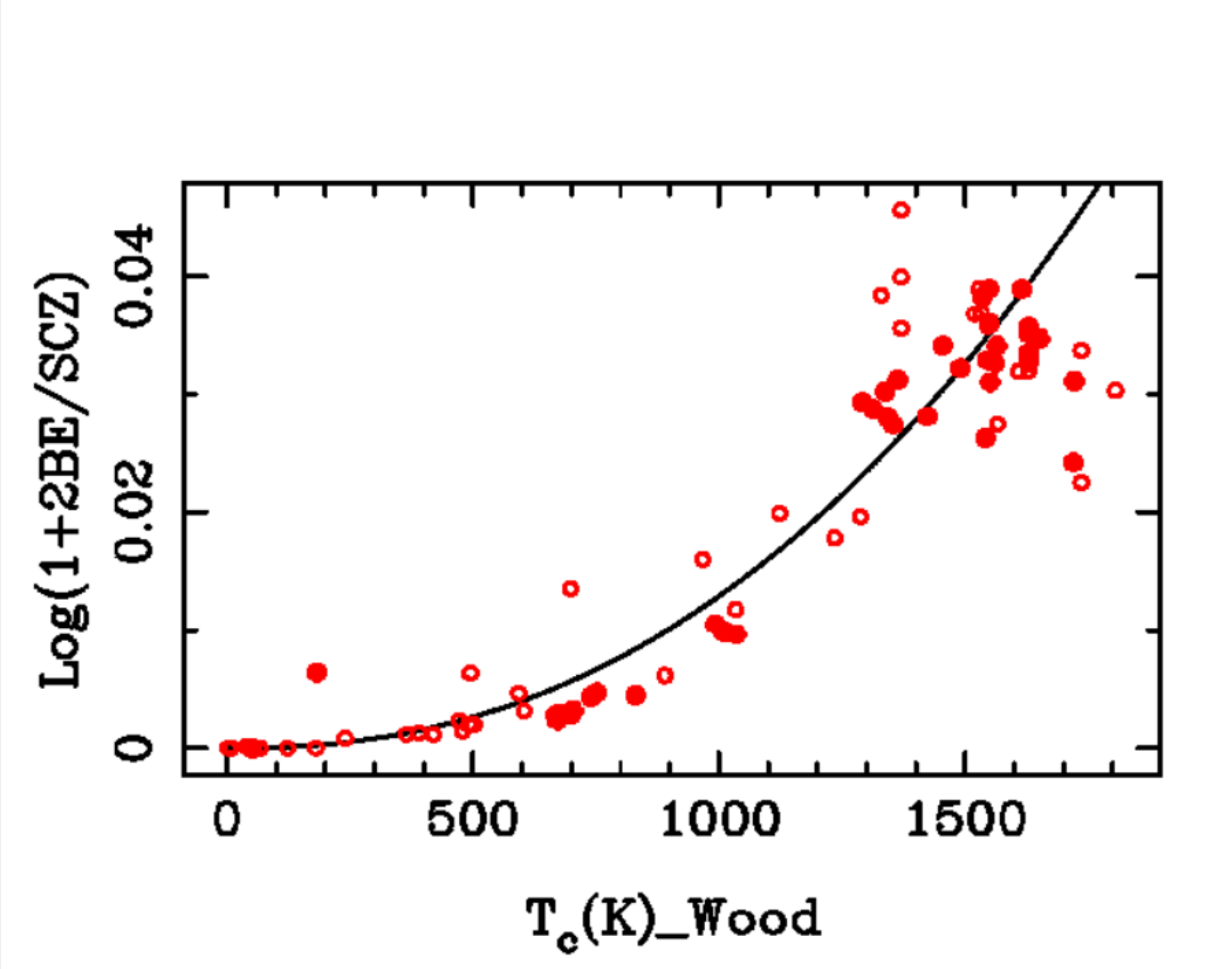}
\caption{Left: Model calculations for the change in abundances
associated with the ingestion of 2 bulk earths from WN18 for the 30 elements
of the BD18 study.
The error bars are from WN18; the largest is for carbon (leftmost point),
whose abundance uncertainty is some 83\%.  Errors for other elements scale proportionally;
the vertical scale does {\em not} apply to the errors.
\newline Right: Same as the left panel, except plotted for all species in WN18.
Points for elements whose solar and/or bulk earth (WN18)
abundances are uncertain by $\ge$ 0.08 dex are designated by small
open circles.  A point for lithium is not plotted. The line is a simple trial and error fit
$\propto 10^{-8.88}\cdot T_{\rm c}^{+2.32}$
 \label{fig:one}}
\end{center}
\end{figure}

In Fig.~\ref{fig:one} (Left), we plot the $change$ in abundance values
for the 30 elements studied in
BD18 for an addition of two earth masses to a solar convection zone.
Many stellar studies consider even fewer elements, and
include  only one or a few neutron-addition elements \citep[e.g.][]{liu20}.
It is difficult to justify a more elaborate fit than a linear one,
because of the
scatter and sparseness of points especially at low $T_{\rm c}$.
Workers typically attempt to fit the points with
a straight line or line segments \citep[cf.][]{liu20,nis20}.

Fig.~\ref{fig:one} (Right) shows all elements from WN18
apart from lithium, which would be a high outlier
at $T_{\rm c} = 1148$K.  It shows a much more detailed picture
than Fig.~\ref{fig:one} (Left).
Using the full data from the WN18, we can explore the
hypothesis of planet (or interplanetary debris) addition
in greater detail when
more stellar data points become available.

Elements with $T_{\rm c} <\sim 500$K are generally depleted
and the logarithm of $(1 + BE/SCZ)$ is
essentially zero.
The slope at low $T_{\rm c}$ would become appreciable only when some 20 bulk earth
masses or more are added.
If the peculiar abundance revealed by PDAs is due to the ingestion
of only one or a few earth-like masses, the slope at small $T_{\rm c}$
should be close to zero.

Oxygen  with $T_{\rm c}= 40$K stands out above the trend for nearby elements
in Fig.~\ref{fig:one} (cf. Tab.~\ref{tab:1})
because much of it is incorporated in
refractory silicate minerals.  That explanation is not relevant
for Pb ($T_{\rm c}=727, 495$K), Ag ($T_{\rm c}=996, 699$K),
Au ($T_{\rm c}=1060, 967$K), and Mn ($T_{\rm c}=1158, 1123$K)
(The first value is from \citet{lod03}, the second from
\citet{wood19}.)  These points follow the
overall trend as high outliers.

The plot for elements with $T_{\rm c} >\sim 1300$K shows a wide
scatter but little evidence of a meaningful trend with $T_{\rm c}$.
Some insight may be gained from plots similar to Fig.~\ref{fig:one} (Right),
but with compositions taken from meteorites (See  Cowley, Bord, and
Y\"{u}ce, AAS237, iPoster 548.11, 2021).  A recent discussion of these
questions is given by \citet{feg20}, who show that much of the scatter at
high $T_{\rm c}$ arises because of systematic abundance differences between
refractory lithophile and siderophile elements (see their Figs. 3-5).




\acknowledgments
We thank N. Grevesse for updated solar abundances, references, and evaluation of their accuracies.
We also thank
W. Clarkson, H. Palme and our referee B. Fegley for useful comments.
This research has made use of the NASA Exoplanet Archive, which is operated
by the California Institute of Technology, under contract with the National Aeronautics
and Space Administration under the Exoplanet Exploration Program.
We also made use of the SIMBAD database,
operated at CDS, Strasbourg, France \citep{wen00}.

\end{document}